\documentclass[aps,preprint,prx,longbibliography]{revtex4-2}  
\usepackage{graphicx,amsmath,subcaption, fullpage}
\usepackage{epsf}
\usepackage{dcolumn}
\usepackage{bm}
\usepackage{subcaption}
\usepackage{amssymb}
\usepackage{array}
\usepackage{varioref}
\usepackage{comment}
\usepackage[normalem]{ulem}
\usepackage{color}
\usepackage[dvipsnames]{xcolor}
\usepackage[colorlinks, linkcolor=blue, citecolor=red]{hyperref}
\usepackage{soul}
\usepackage{float}
\usepackage{url} 

\restylefloat{table}

\usepackage{babel}
\usepackage{natbib}

\begin{document}

\title{
RKKY interaction mediated by a spin-polarized 2D electron gas with Rashba and altermagnetic coupling
}

\author{Anirban Kundu}
\affiliation{Asia Pacific Center for Theoretical Physics, POSTECH, Pohang 37673, Korea.}


\begin{abstract}
Magnetic interactions between impurity spins play a crucial role in determining magnetic configurations in spintronic systems.
Using a Green’s function formalism, we investigate Ruderman–Kittel–Kasuya–Yosida (RKKY) exchange interactions between two localized spins mediated by a two-dimensional electron gas arising from two spin-polarized bands with Rashba spin–orbit coupling (RSOC) and altermagnetic dispersion. 
We analyze two distinct regimes: (a) strong out-of-plane ferromagnetic order, and (b) weak in-plane order. 
For out-of-plane magnetization, the Heisenberg, Ising, and Dzyaloshinskii–Moriya (DM) exchange interaction terms exhibit oscillatory spatial modulations and asymptotically decay as $1/R^{2}$ with impurity separation. 
Besides, all exchange terms display beating-like patterns that can be tuned via the exchange coupling strength between conduction electrons and ferromagnetic ordering. 
The DM vector lies within the two-dimensional plane, with the DM interaction being odd in the RSOC strength, oscillatory, and increasing in magnitude with RSOC strength. 
In contrast, it is even in the out-of-plane Zeeman field strength and oscillatory. 
Furthermore, the Heisenberg interaction exhibits a non-linear dependence on the altermagnetic band parameter.
In the case of weak in-plane order, the Heisenberg exchange interaction shows a non-linear dependence on the in-plane exchange coupling strength. 

\end{abstract}

\maketitle

\section{Introduction}
The Ruderman-Kittel-Kasuya-Yosida (RKKY) interaction  is an indirect exchange coupling mechanism between localized magnetic  impurities  mediated by conduction electrons in a host material. 
This indirect exchange interaction  arises as a consequence of the perturbative energy correction due to the  coupling between the conduction electrons and localized impurity spins. 
The RKKY interaction has been extensively studied in various condensed matter systems, including 
dilute magnetic alloys \cite{Fert1981,Dietl2000}, 
rare-earth compounds \cite{Stewart1984}, 
semiconductors \cite{PhysRevLett.44.1538,  Kogan2004, Craig2004, Valizadeh2015,Wang2017,Bouaziz2017}, 
graphene~\cite{PhysRevB.76.184430, Bunder2009, Sherafati2011, Power2013, gorman2013rkky, PhysRevB.87.045422, wkp4-2n5b}, 
topological insulators~\cite{ Liu2009, Zhu2011},
and in Dirac and Weyl semimetals \cite{Hosseini2015, Chang2015,PhysRevB.99.165302,PhysRevB.102.165110, PhysRevB.107.165147,PhysRevB.108.134428}
 since its original theoretical derivation \cite{Ruderman1954, Kasuya1956, Yosida1957}.
The interaction significantly depends on the dimensionaly, and a number of studies has been made to addess its role \cite{PhysRevB.105.094427,PhysRevB.69.121303,PhysRevB.58.3584, LI2022169607, PhysRevB.69.121303}. 
Besides, the interaction depends on the dispersion of the conduction electrons. 
For a spin-degenerate band, the RKKY interaction favors collinear alignment and exhibits an oscillatory decay with impurity separation, alternating between ferromagnetic and antiferromagnetic alignments.
 However, the presence of momentum dependent spin-orbit coupling leads to additional terms, among which a crucial one is Dzyaloshinskii--Moriya  (DM)  exchange interaction.



Recently, alongside ferromagnetism and antiferromagnetism, altermagnetism has emerged as a third fundamental class of magnetism, attracting rapidly growing theoretical and experimental interest. 
This phase is characterized by a compensated magnetic order, akin to antiferromagnets, yet it exhibits a strong momentum-dependent spin splitting in the electronic band structure \cite{smejkal2022emerging, 00000202409327}.
The $d$-wave ($k_x^2 - k_y^2$) symmetry of this spin-split band and its profound implications for electron transport are established in the foundational theoretical work of Ref. \cite{smejkal2022beyond}.
Experimentally, such spin-split band has been confirmed recently in chromium antimonide (CrSb), using spin-polarized angle-resolved photoemission spectroscopy \cite{Reimers2024}.
The experimental confirmations were reported in MnTe and RuO$_2$ systems~\cite{Krempasky2024_Nature,Fedchenko2024_SciAdv,Amin2024,plouff2024revisitingaltermagnetismruo2study}.
Besides, a significant efforts have been met to realize the spin-polarize current \cite{smejkal2022beyond}. 

%
Currently, the effect of altermagnetic band structure to the  RKKY interaction in the weak RSOC limit  has been studied by Ref. \cite{PhysRevB.110.054427}.
A  separate study by Ref. \cite{Yarmohammadi_2025} demonstrated that the RKKY interaction in altermagnets can be tuned using circularly polarized light. Through the inverse Faraday effect, the light induces an effective anisotropic term in the Floquet Hamiltonian, which, in combination with spin-orbit coupling, yields the desired modification of the interaction in the limit of weak RSOC.
Both the previous theoretical works considered 2D magnetic systems with weak RSOC for their model Hamiltonian.

However,  for a 2D magnetic system,  strong RSOC is often a prerequisite  to observe novel spin-dependent phenomena in the domain of spintronics research.
Furthermore, a theoretical description of the RKKY interaction across the full range of RSOC strengths remains lacking, representing a significant gap in the literature.
To address this issue, we develop a general model for the RKKY interaction in an altermagnetic Rashba system without restriction on the strength of the RSOC.
In particular, we consider a ferromagnetic system when RSOC is stronger than altermagnetic band parameter.
Considering the broken inversion symmetry, we systematically investigate all contributions to the RKKY exchange interaction within a single-particle Green’s function formalism. 
In particular, we analyze two distinct cases:
	(a) ferromagnetic order oriented out-of-plane, relevant for systems with strong perpendicular magnetic anisotropy, and
	(b) ferromagnetic order confined to the in-plane direction.


\section{Method}

We consider a general model Hamiltonian for a two-dimensional electron gas (2DEG)  in the presence
of a ferromagnetic order, RSOC, and altermagnetic coupling,
\begin{align}
H_{\boldsymbol{k}} & =\epsilon_{\boldsymbol{k}}\sigma_{0}+\left(\alpha\hat{z}\times\boldsymbol{k}+J_{\parallel}\boldsymbol{m}_{\parallel}\right)\cdot\boldsymbol{\sigma}+J_{\perp}\boldsymbol{m}_{\perp}\sigma_{3}+\beta\left(k_{x}^{2}-k_{y}^{2}\right)\sigma_{3}\label{eq: main_hamiltonian}
\end{align}
where, $\epsilon_{\bm{k}}=\hbar^{2}k^{2}/2m$, $\alpha$ is the RSOC strength, $\beta$ is the altermagnetic coupling
parameter, $\sigma_{i}$'s ($\forall i\in[1,3]$) are the Pauli matrices
and $\sigma_{0}$ is the identity matrix;  $\boldsymbol{m}_{\parallel}=m_{x}\hat{x}+m_{y}\hat{y}$
is the in-plane and $\boldsymbol{m}_{\perp}=m_{z}\hat{z}$ is the
out-of-plane component of magnetic moment vector, $J_{\perp}$, $J_{\parallel}$
are the perpendicular and in-plane exchange coupling strengths. 
The $s$-$d$ interaction between the conduction electrons and  two impurity-spins at positions $\boldsymbol{ R}_1$ and $\boldsymbol{ R}_2$  is, 
\begin{align}
H^{int}=\mathcal{J}\left[\boldsymbol{\sigma}\cdot\boldsymbol{S}_{1}\delta\left(\boldsymbol{r}-\boldsymbol{R}_{1}\right)+\boldsymbol{\sigma}\cdot\boldsymbol{S}_{2}\delta\left(\boldsymbol{r}-\boldsymbol{R}_{2}\right)\right].
\end{align}
Treating the above interaction as a perturbation to the Hamiltonian in Eq. (\ref{eq: main_hamiltonian}), the second-order correction gives rise to
the RKKY interaction between a pair of spins $S_{1}$ and $S_{2}$,
and is given by \cite{mahan2000many}, 
\begin{align}
H_{\text{RKKY}} & =-\frac{\mathcal{J}^{2}}{\pi}\text{Im}\int_{-\infty}^{\epsilon_{F}}d\epsilon\text{ Tr}\left[\left(S_{1}\cdot\sigma\right)G\left(\boldsymbol{R},\epsilon\right)\left(S_{2}\cdot\sigma\right)G\left(-\boldsymbol{R},\epsilon\right)\right]\label{eq: rkky-main-expression}.
\end{align}
The Green's function corresponding to the Hamiltonian in Eq.  (\ref{eq: main_hamiltonian}) is calculated using $G\left(\boldsymbol{k},\epsilon\right)=\left(\epsilon-H_{\boldsymbol{k}}+i\eta\right)^{-1}$
with $\eta\rightarrow0$, and  has the following expression, 
\begin{align}
G\left(\boldsymbol{k},\epsilon\right) & =D_{\boldsymbol{k}}^{-1}\left[\left(\epsilon-\epsilon_{\boldsymbol{k}}\right)\sigma_{0}+\left(\alpha k_{y}+m_{x}\right)\sigma_{1}+\left(-\alpha k_{x}+m_{y}\right)\sigma_{2}+\left(J_{\perp}m_{z}+\beta_{\boldsymbol{k}}\right)\sigma_{3}\right].
\label{eq: gk-1}
\end{align}
With the aim to essentially capture the physics of RKKY exchange interaction in the presence of a spin-polarized band with strong Rashba coupling, 
we assume, $J_{\parallel}\text{, }\beta\ll J_{\perp}\text{, }\alpha\text{, }\epsilon_{F}$,
and $D_{\boldsymbol{k}}=\left(\epsilon-\epsilon_{\boldsymbol{k}}+i\delta\right)^{2}-\left(\alpha^{2}k^{2}+J_{\perp}^{2}m_{z}^{2}\right)$.
Please note that, for $J_{\perp}=0$, i.e., in the absence of out-of-plane ferromagnetic order, the expansion is still
valid.
 Expressing the Green's function in Eq. (\ref{eq: gk-1}) in the Pauli matrix basis yields,
\begin{equation}
G\left(\boldsymbol{k},\epsilon\right)=G_{0}\left(\boldsymbol{k},\epsilon\right)\sigma_{0}+\sum_{i}G_{i}\left(\boldsymbol{k},\epsilon\right)\sigma_{i}, 
\label{eq: gr_k}
\end{equation}
with $i\in [1,3]$, and 
\begin{align*}
G_{0}\left(\boldsymbol{k},\epsilon\right) & =\left(\epsilon-\epsilon_{\boldsymbol{k}}\right)D_{\boldsymbol{k}}^{-1}\\
G_{1}\left(\boldsymbol{k},\epsilon\right) & =\left(\alpha k_{y}+m_{x}\right)D_{\boldsymbol{k}}^{-1}\\
G_{2}\left(\boldsymbol{k},\epsilon\right) & =\left(-\alpha k_{x}+m_{y}\right)D_{\boldsymbol{k}}^{-1}\\
G_{3}\left(\boldsymbol{k},\epsilon\right) & =\left(J_{\perp}m_{z}+\beta_{\boldsymbol{k}}.\right)D_{\boldsymbol{k}}^{-1}.
\end{align*}
The real-space Green's function is obtained via the Fourier transform, 
\begin{equation}
G\left(\boldsymbol{R},\epsilon\right)=\int d\boldsymbol{k}~e^{i\boldsymbol{k}\cdot\boldsymbol{R}}G\left(\boldsymbol{k},\epsilon\right)=G_{0}\left(\boldsymbol{R},\epsilon\right)\sigma_{0}+\sum_{i}G_{i}\left(\boldsymbol{R},\epsilon\right)\sigma_{i},
\label{eq: gr_k}
\end{equation}
with $i\in [1,3]$, and 
\begin{align}
G_{0}\left(\boldsymbol{R},\epsilon\right) & =\left(\epsilon+\frac{\hbar^{2}}{2m}\frac{\partial^2}{\partial R^{2}}\right)I\left(R,\epsilon\right)\nonumber \\
G_{1}\left(\boldsymbol{R},\epsilon\right) & =\left(-i\alpha\frac{\partial}{\partial R_{y}}+m_{x}\right)I\left(R,\epsilon\right)\nonumber \\
G_{2}\left(\boldsymbol{R},\epsilon\right) & =\left(i\alpha\frac{\partial}{\partial R_{x}}+m_{y}\right)I\left(R,\epsilon\right)\nonumber \\
G_{3}\left(\boldsymbol{R},\epsilon\right) & =\left(J_{\perp}m_{z}-\beta\left(\frac{\partial^2}{\partial R_{x}^{2}}-\frac{\partial^2}{\partial R_{y}^{2}}\right)\right)I\left(R,\epsilon\right)\label{eq: gr}
\end{align}
We note that the components $G_1$, $G_2$, and $G_3$ are non-zero solely due to the presence of spin-split bands. The corresponding integral is given by
\begin{equation}
I\left(R,\epsilon\right)=\int \frac{d\boldsymbol{k}}{\left( 2\pi\right)^2} e^{i\boldsymbol{k}\cdot\boldsymbol{R}}D_{\boldsymbol{k}}^{-1}
=\frac{2m^{2}}{\pi\hbar^{4}\left(k_{+}^{2}-k_{-}^{2}\right)}\left(k_{+}H_{0}^{(1)}(k_{+}R)-k_{-}H_{0}^{(1)}(k_{-}R)\right)
\end{equation}
with, {\scriptsize $ k_{\pm}=\left(k_{F}/\sqrt{2}\right)\sqrt{\left(\alpha k_{F}/\epsilon_{F}\right)^{2}+2\left(\epsilon/\epsilon_{F}\right)\pm\sqrt{\left(\alpha k_{F}/\epsilon_{F}\right)^{4}+4\left(\epsilon/\epsilon_{F}\right)\left(\alpha k_{F}/\epsilon_{F}\right)^{2}+4\left(J_{\perp}^{2}m_{z}^{2}/\epsilon_{F}^{2}\right)}}+i\delta$}{\tiny{}
}
and the Fermi energy $\epsilon_{F}=\hbar^{2}k_{F}^{2}/2m$  (see Supplemental Eq. (S4) for details). Since the RKKY exchange
is intrinsically long-ranged, we examine its behavior when the separation
between the impurities is large, i.e., $k_{F}R\gg1$. In this limit, the asymptotic expansion of the Hankel function is, $H_{0}^{(1)}(kR)\simeq\sqrt{\frac{2}{\pi kR}}\,e^{i(kR-\frac{\pi}{4})}$. Accordingly, the radial integral $I(R,\epsilon)$
becomes, 
\begin{equation}
I\left(R,\epsilon\right)=\frac{2m^2}{\pi \hbar^{4}\left(k_{+}^{2}-k_{-}^{2}\right)}e^{-i\frac{\pi}{4}}\sqrt{\frac{2}{\pi R}}\left(\sqrt{k_{+}}\,e^{ik_{+}R}-\sqrt{k_{-}}\,e^{ik_{-}R}\right).
\label{eq: ir}
\end{equation}
The above expression represents the superposition of two waves with different frequencies, which is expected to generate beating patterns.
Now putting  Eq. (\ref{eq: ir}) into Eq. ({\ref{eq: rkky-main-expression}}), the final form of the RKKY interaction becomes,
\begin{align}
H_{\text{RKKY}} & =2\sum_{i}J_{ii}\left(\boldsymbol{R}\right)S_{1i}S_{2i}+\boldsymbol{D}\left(\boldsymbol{R}\right)\cdot\boldsymbol{S}_{1}\times\boldsymbol{S}_{2}+\sum_{i\neq j}J_{ij}\left(\boldsymbol{R}\right)\left(S_{1i}S_{2j}+S_{1j}S_{2i}\right)
\end{align}
where, $i,j \in {x,y,z}$, and  
\begin{widetext}  
{\small
\begin{align}
J_{xx}\left(\boldsymbol{R}\right) & =-\frac{\mathcal{J}^{2}}{\pi}\text{Im}\int_{-\infty}^{\epsilon_{F}}d\epsilon\left[G_{0}^{2}\left(R,\epsilon\right)+G_{1}\left(\boldsymbol{R},\epsilon\right)G_{1}\left(-\boldsymbol{R},\epsilon\right)-G_{2}\left(\boldsymbol{R},\epsilon\right)G_{2}\left(-\boldsymbol{R},\epsilon\right)-G_{3}\left(\boldsymbol{R},\epsilon\right)G_{3}\left(-\boldsymbol{R},\epsilon\right)\right]\nonumber \\
J_{yy}\left(\boldsymbol{R}\right) & =-\frac{\mathcal{J}^{2}}{\pi}\text{Im}\int_{-\infty}^{\epsilon_{F}}d\epsilon\left[G_{0}^{2}\left(R,\epsilon\right)-G_{1}\left(\boldsymbol{R},\epsilon\right)G_{1}\left(-\boldsymbol{R},\epsilon\right)+G_{2}\left(\boldsymbol{R},\epsilon\right)G_{2}\left(-\boldsymbol{R},\epsilon\right)-G_{3}\left(\boldsymbol{R},\epsilon\right)G_{3}\left(-\boldsymbol{R},\epsilon\right)\right]\nonumber \\
J_{zz}\left(\boldsymbol{R}\right) & =-\frac{\mathcal{J}^{2}}{\pi}\text{Im}\int_{-\infty}^{\epsilon_{F}}d\epsilon\left[G_{0}^{2}\left(R,\epsilon\right)-G_{1}\left(\boldsymbol{R},\epsilon\right)G_{1}\left(-\boldsymbol{R},\epsilon\right)-G_{2}\left(\boldsymbol{R},\epsilon\right)G_{2}\left(-\boldsymbol{R},\epsilon\right)+G_{3}\left(\boldsymbol{R},\epsilon\right)G_{3}\left(-\boldsymbol{R},\epsilon\right)\right]\nonumber \\
D_{x}\left(\boldsymbol{R}\right) & =-\frac{\mathcal{J}^{2}}{\pi}\text{Im}\int_{-\infty}^{\epsilon_{F}}d\epsilon\text{ }\left(-i\right)\text{ }G_{0}\left(R,\epsilon\right)\left[G_{1}\left(\boldsymbol{R},\epsilon\right)-G_{1}\left(-\boldsymbol{R},\epsilon\right)\right]\nonumber \\
D_{y}\left(\boldsymbol{R}\right) & =-\frac{\mathcal{J}^{2}}{\pi}\text{Im}\int_{-\infty}^{\epsilon_{F}}d\epsilon\text{ }\left(-i\right)G_{0}\left(R,\epsilon\right)\left[G_{2}\left(\boldsymbol{R},\epsilon\right)-G_{2}\left(-\boldsymbol{R},\epsilon\right)\right]\nonumber \\
D_{z}\left(\boldsymbol{R}\right) & =0\nonumber \\
J_{xy}\left(\boldsymbol{R},\epsilon\right) & =-\frac{\mathcal{J}^{2}}{\pi}\text{Im}\int_{-\infty}^{\epsilon_{F}}d\epsilon\text{ }\left[G_{1}\left(\boldsymbol{R},\epsilon\right)G_{2}\left(-\boldsymbol{R},\epsilon\right)+G_{1}\left(-\boldsymbol{R},\epsilon\right)G_{2}\left(\boldsymbol{R},\epsilon\right)\right]\nonumber \\
J_{yz}\left(\boldsymbol{R},\epsilon\right) & =-\frac{\mathcal{J}^{2}}{\pi}\text{Im}\int_{-\infty}^{\epsilon_{F}}d\epsilon\text{ }\left[G_{2}\left(\boldsymbol{R},\epsilon\right)G_{3}\left(-\boldsymbol{R},\epsilon\right)+G_{2}\left(-\boldsymbol{R},\epsilon\right)G_{3}\left(\boldsymbol{R},\epsilon\right)\right]\nonumber \\
J_{zx}\left(\boldsymbol{R},\epsilon\right) & =-\frac{\mathcal{J}^{2}}{\pi}\text{Im}\int_{-\infty}^{\epsilon_{F}}d\epsilon\text{ }\left[G_{1}\left(\boldsymbol{R},\epsilon\right)G_{3}\left(-\boldsymbol{R},\epsilon\right)+G_{1}\left(-\boldsymbol{R},\epsilon\right)G_{3}\left(\boldsymbol{R},\epsilon\right)\right]\label{eq: jii}
\end{align}
}
\end{widetext}
Here, $J_{xx}$, $J_{yy}$ are the in-plane Hisenberg interaction terms, $J_{zz}$ is Ising interaction term, and ${D}_i(\boldsymbol{R})$ denotes components of the DM vector. 
We note that Heisenberg and and Ising terms has been calculated recently by Ref. \cite{Lee_2025} using a solely altermagnetic band.
For our two-dimensional system, the out-of-plane component vanishes, i.e., $D_z(\boldsymbol{R}) = 0$. We note that a finite DM interaction arises exclusively from the presence of RSOC. For details see Eqs. (S7) and (S8) in the Supplemental Information (SI).


\section{Results}
We discuss the application of our derived methods for two types of ferromagnetic configuration, out-of-plane and in-plane. 

\subsection{Altermagnetic effects: out-of-plane ferromagntic order }   
In the absence of an in plane spin-orientation, i.e., $ \boldsymbol{m}_{\parallel}=0$, the
Hamiltonian in Eq. (\ref{eq: main_hamiltonian}), represents
two spin-split bands of a 2D ferromagnet with out-of-plane orientation
of parallel spins along with combined RSOC and altermagnetic
effects as follows, 
\begin{equation}
H\left(\boldsymbol{k}\right)=\epsilon_{\boldsymbol{k}}\sigma_{0}+\alpha\boldsymbol{\sigma}\cdot\hat{z}\times\boldsymbol{k}+J_{\perp}\sigma_{3}+\beta\left(k_{x}^{2}-k_{y}^{2}\right)\sigma_{3},
\end{equation}
where, we have set $\left|\boldsymbol{m}_{\perp}\right|=1$
without loss of  generality.
 To calculate the exchange interaction due to the above Hamiltonian
  we
set $\boldsymbol{m}_{\parallel}=0$ in Eq. (\ref{eq: jii})
and numerically carry out the energy integration.



\begin{figure*}
    \centering
    
    \begin{minipage}[t]{0.49\hsize}  
        \centering
        \includegraphics[width=\hsize, height=0.3\textheight, keepaspectratio]{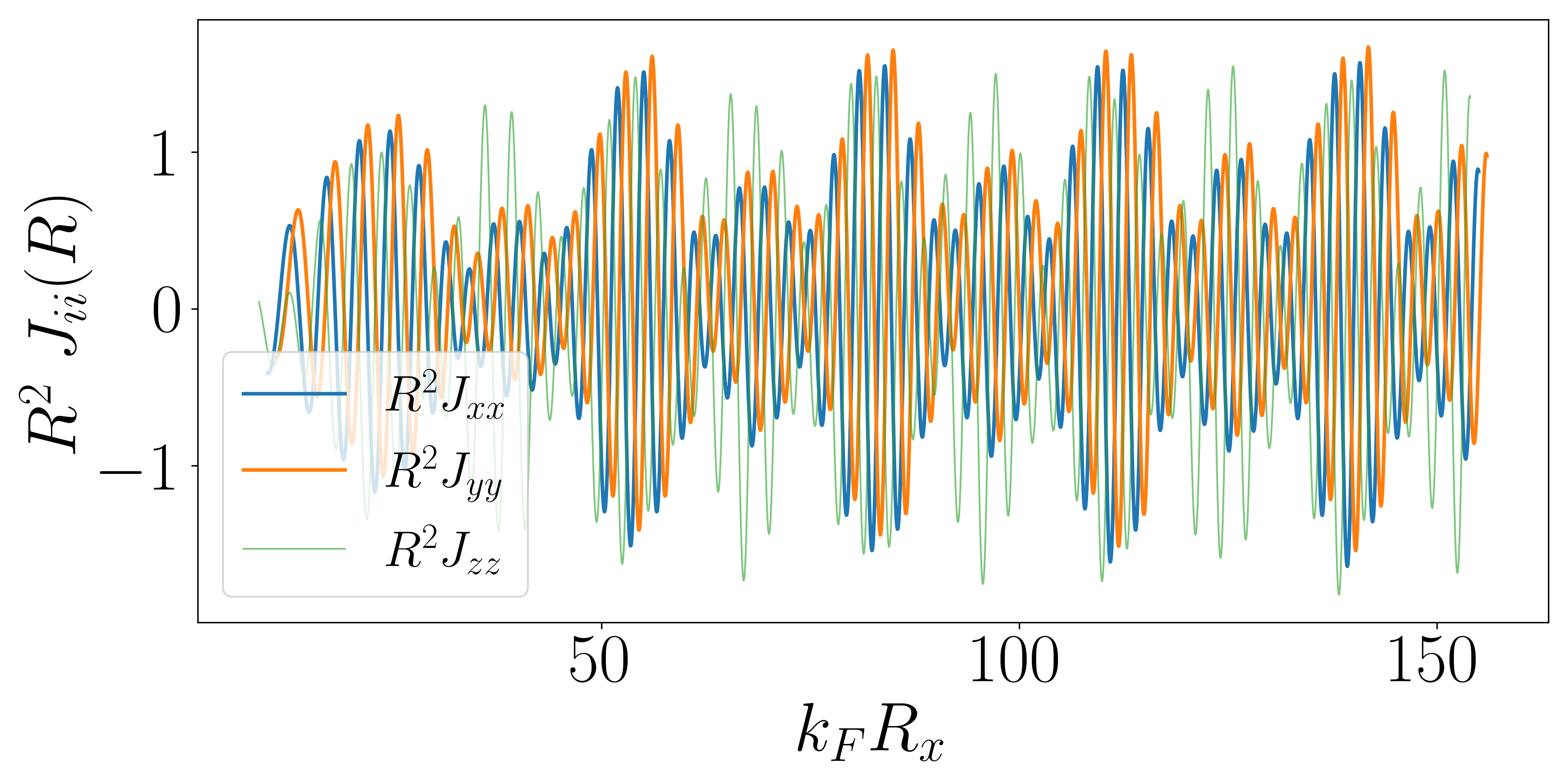}
        \subcaption{ }
    \end{minipage}%
    \hfill
    \begin{minipage}[t]{0.49\hsize}
        \centering
        \includegraphics[width=\hsize, height=0.3\textheight, keepaspectratio]{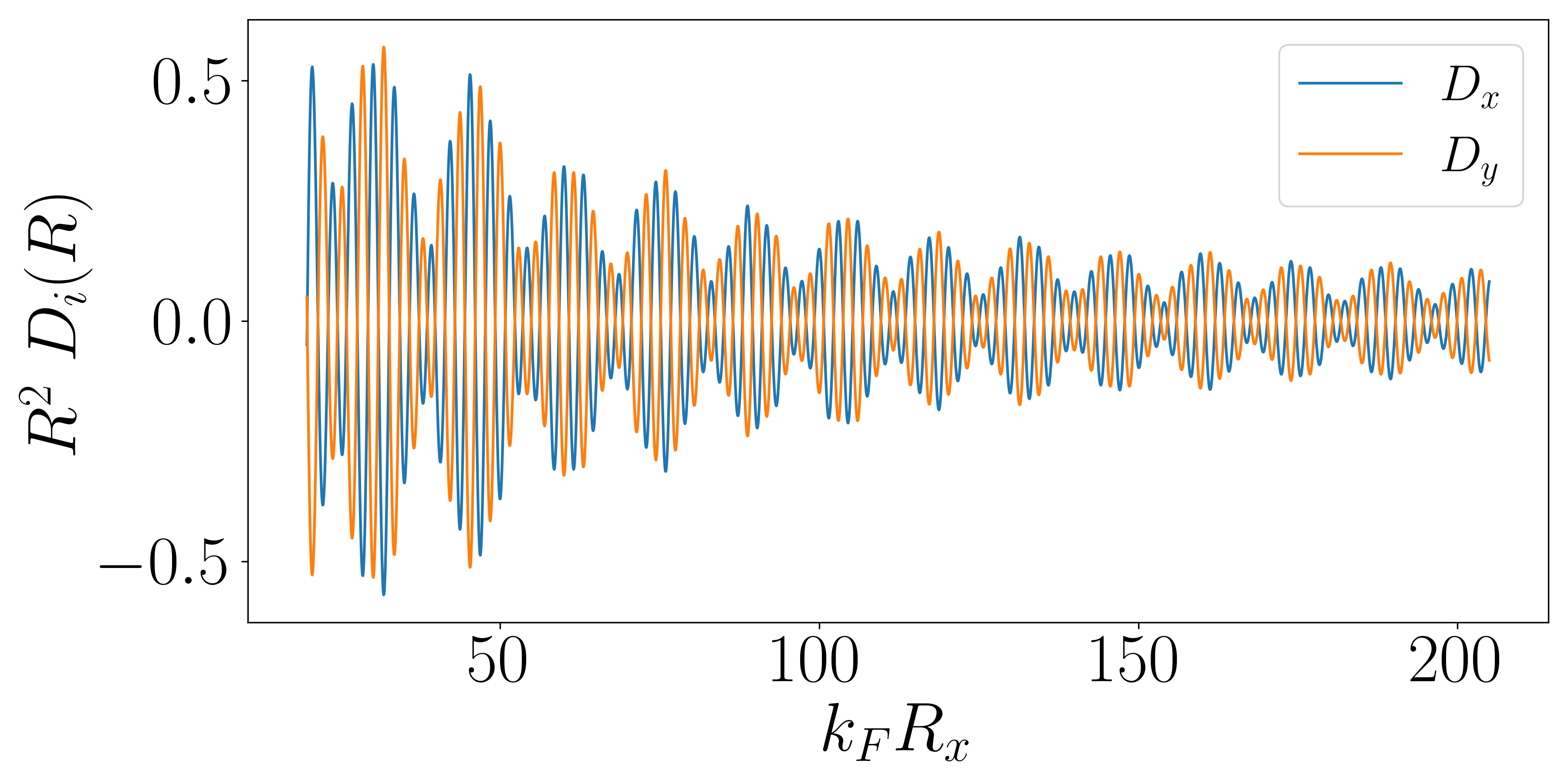}
        \subcaption{}
    \end{minipage}

    
    \begin{minipage}[t]{0.49\hsize}
        \centering
        \includegraphics[width=\hsize, height=0.3\textheight, keepaspectratio]{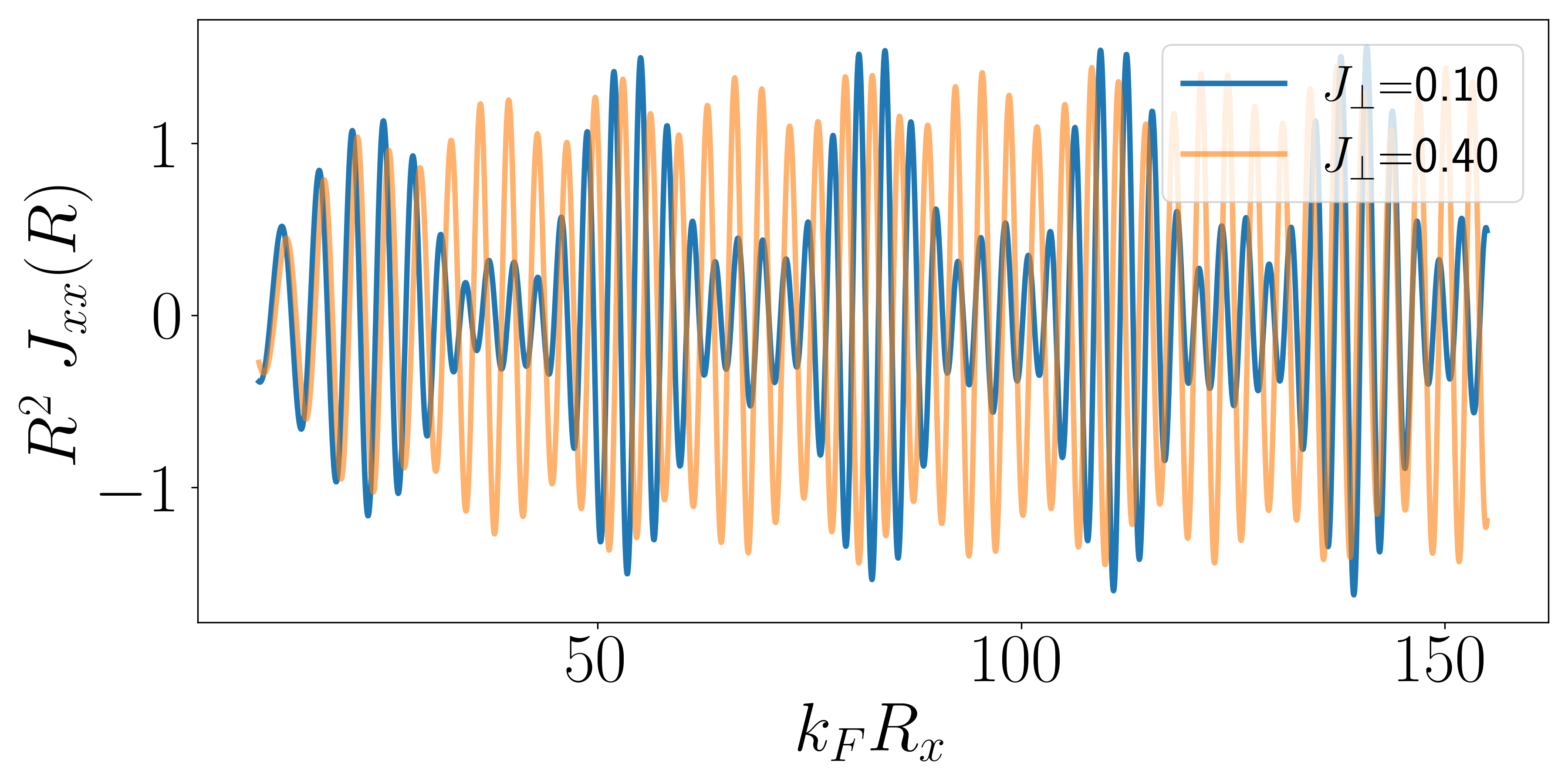}
        \subcaption{}
    \end{minipage}%
    \hfill
    \begin{minipage}[t]{0.49\hsize}
        \centering
        \includegraphics[width=\hsize, height=0.3\textheight, keepaspectratio]{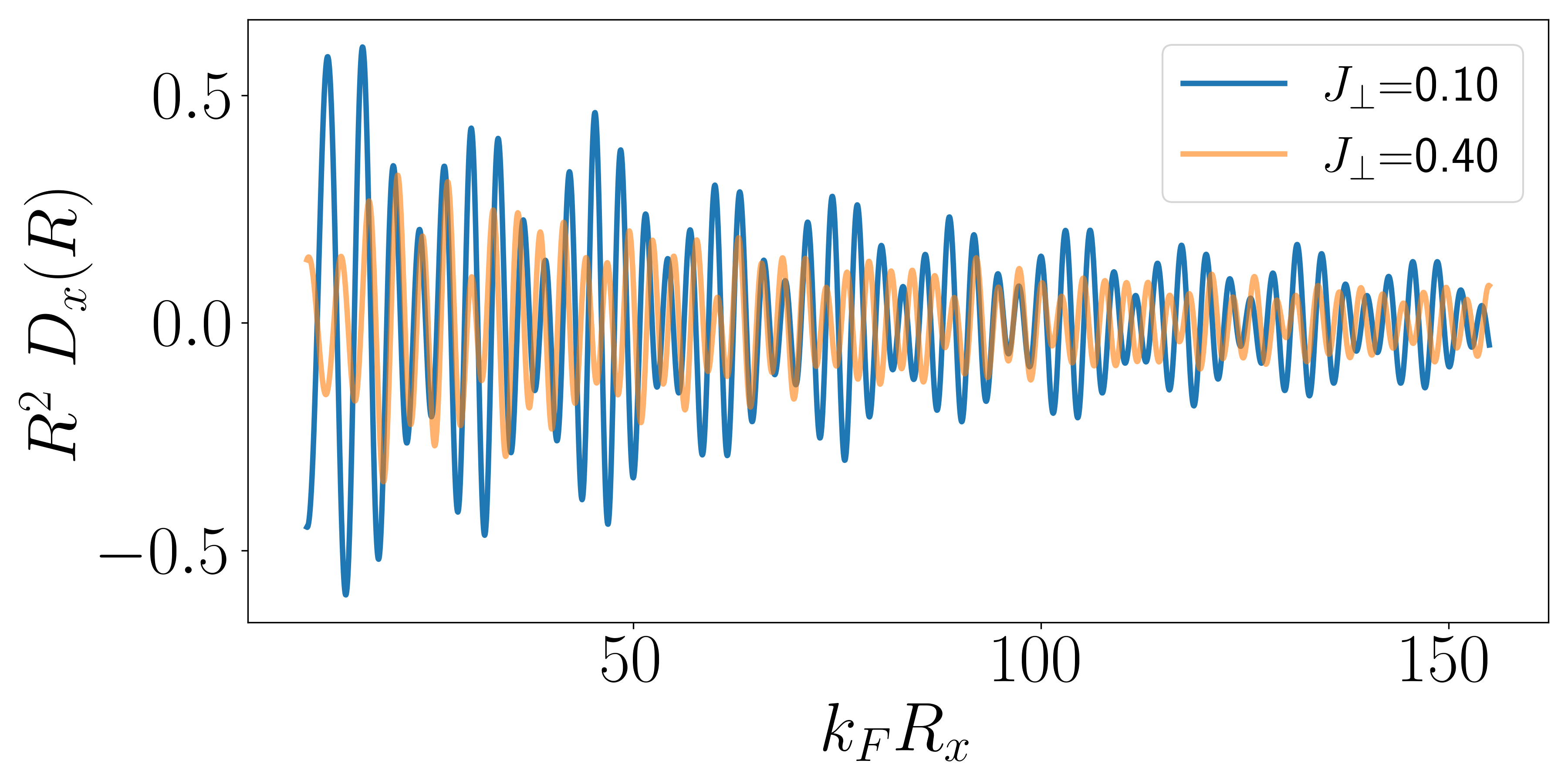}
        \subcaption{}
    \end{minipage}    
    \caption{(a), (b), Variation of $J_{ii}$ and $D_x$ with separation distance $k_F R$ with specific parameters $\alpha/\epsilon_F = 0.2$, $\beta /\epsilon_F= 0.05$ and $J_\perp/\epsilon_F = 0.1$. 
(c), (d), Effect of $J_\perp$  on the  exchange interaction with the same set of parameteres. 
For all plots $y$-axis is scaled in arbitrary units (a.u.).
}
\label{fig-R-variation}
\end{figure*}



In Fig.~\ref{fig-R-variation}(a), we plot the exchange interaction terms
$R^{2}J_{xx}(\boldsymbol{R})$, $R^{2}J_{yy}(\boldsymbol{R})$, and
$R^{2}J_{zz}(\boldsymbol{R})$ as functions of $k_FR_{x}$ (with fixed $k_FR_{y}=15$). 
Now we obtain, from Eq. (\ref{eq: gr}) that $\mathcal{R}(\pi/2)G_{1(2)}(\boldsymbol{R}, \epsilon)=G_{2(1)}(\boldsymbol{R}, \epsilon)$ where $\mathcal{R}$ is the rotation operator in 2D coordinate space.  
As a consequence, the behavior of $J_{yy}(\boldsymbol{R})$ is obtained
by a rotation of the axes by an angle $\pi/2$.
 When the distance of separation is large, the envelope
peaks saturate to constant values, confirming the expected $1/R^{2}$
decay for a 2D magnetic system. Besides, the
exchange interaction terms display a spatial beating pattern, a phenomenon that
has also been observed in systems governed by spin-split Hamiltonians.
Such beating arises from the coexistence of two Fermi wave vectors
with distinct magnitudes at the Fermi surface. The nodal amplitudes
of the envelope functions are governed by two parameters, the RSOC strength $\alpha$ and exchange coupling strength $J_{\perp}$. An increase in $J_{\perp}$
leads to a broadening of the envelope amplitude at the node, suppressing
the beating effect as shown in Fig.~\ref{fig-R-variation}(c). In contrast,
for larger $\alpha$, the beating disappears entirely and is replaced
by two distinct envelopes in the spatial variation. 

Fig.~\ref{fig-R-variation}(b), displays the spatial dependence of the in-plane
DMI components, plotted as $R^{2}D_{x(y)}(\boldsymbol{R})$ versus $k_FR_{x}$ while we keep fixed $k_FR_{y}=50$ . Similar
to the Heisenberg terms, the DM vector components exhibit a characteristic
$1/R^{2}$ decay at large distances. The DM interaction strength shows significant
dependence on the parameters $\alpha$, $J_{\perp}$, and $\beta$.

\begin{figure}
    \centering
    
    \begin{minipage}[t]{0.45\hsize}  
        \centering
        \includegraphics[width=\hsize,]{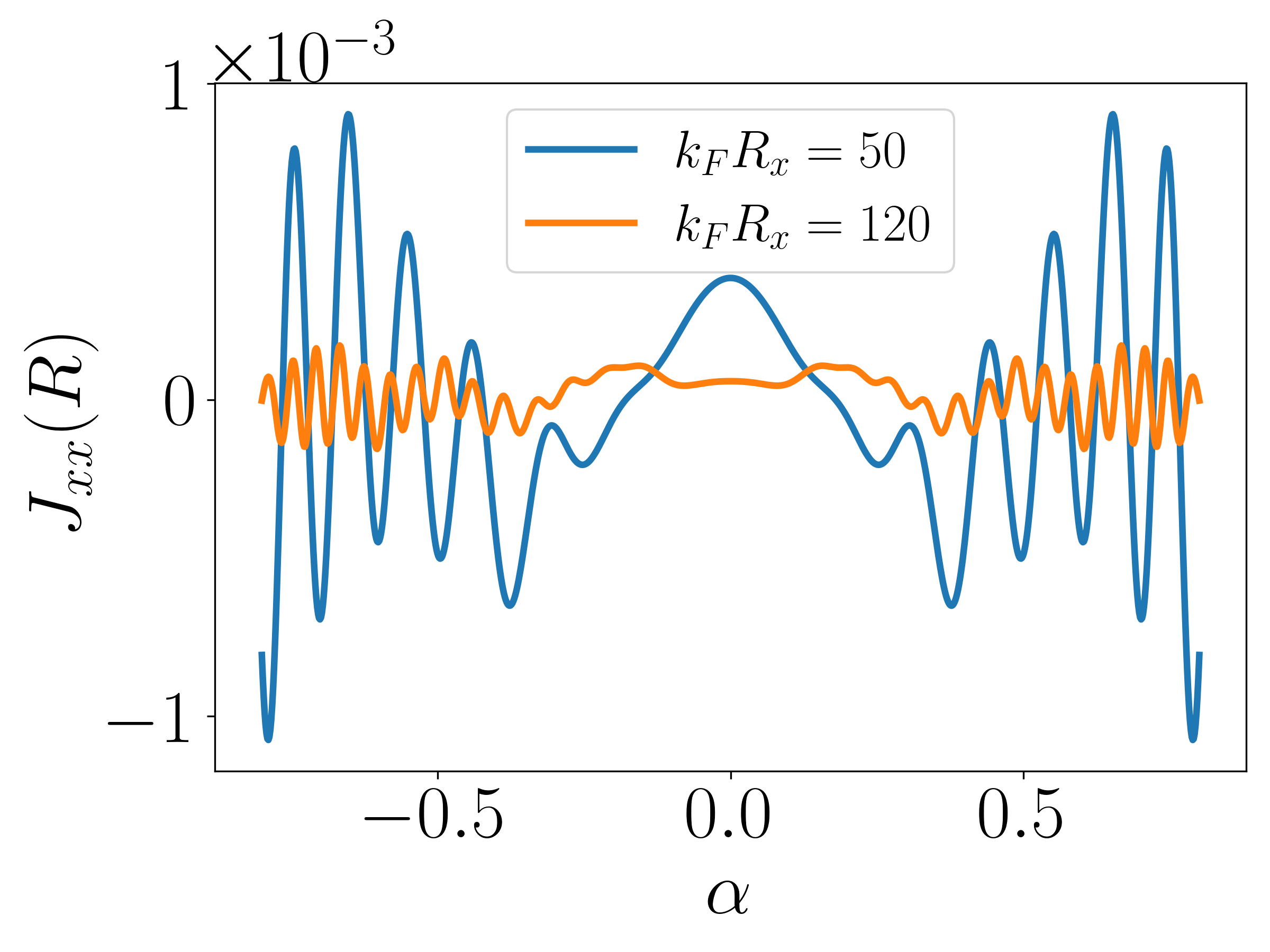}
        \subcaption{ }
    \end{minipage}%
    \hfill
    \begin{minipage}[t]{0.45\hsize}
        \centering
        \includegraphics[width=\hsize]{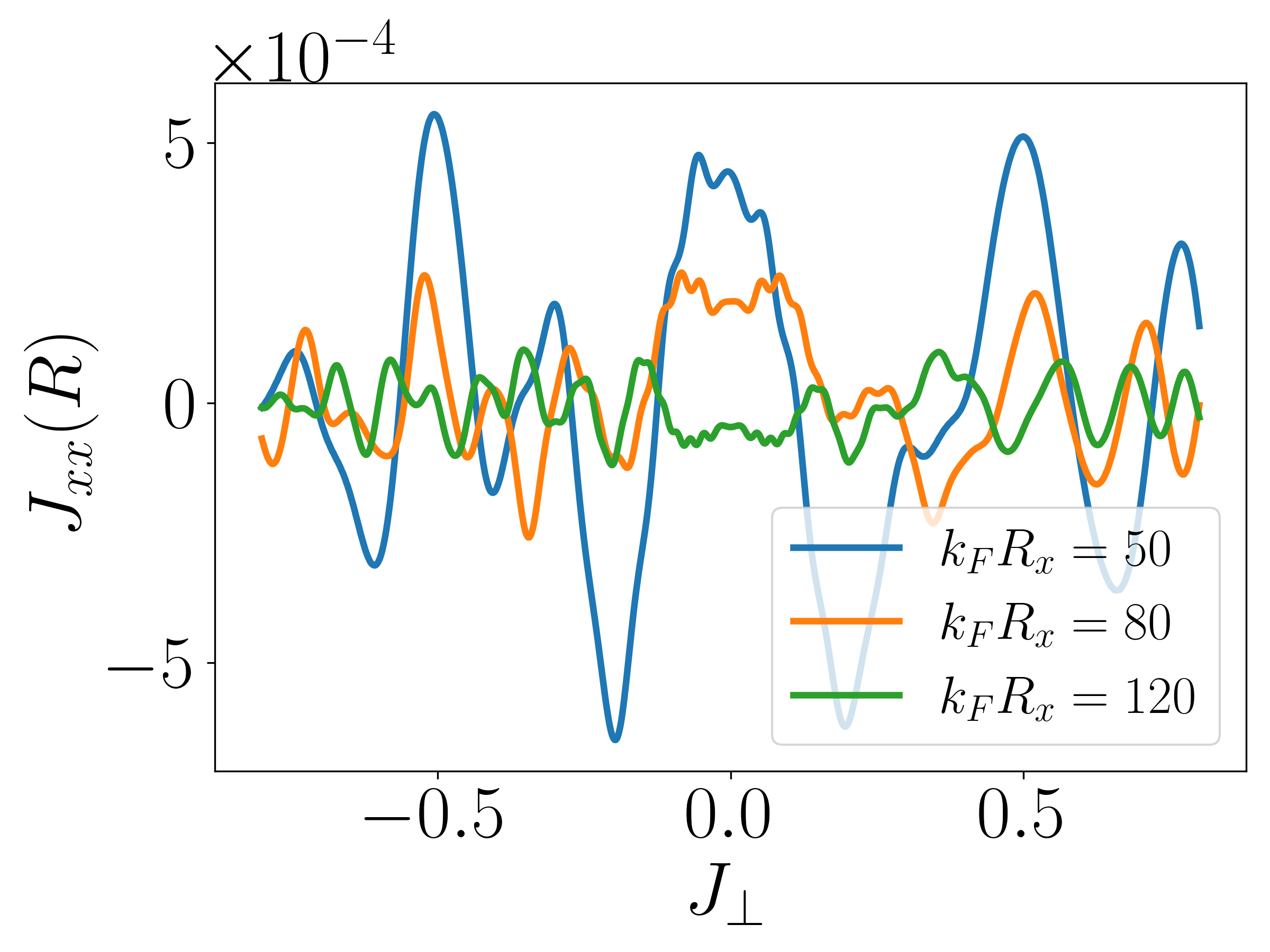}
        \subcaption{}
    \end{minipage}

    
    \begin{minipage}[t]{0.45\hsize}
        \centering
        \includegraphics[width=\hsize]{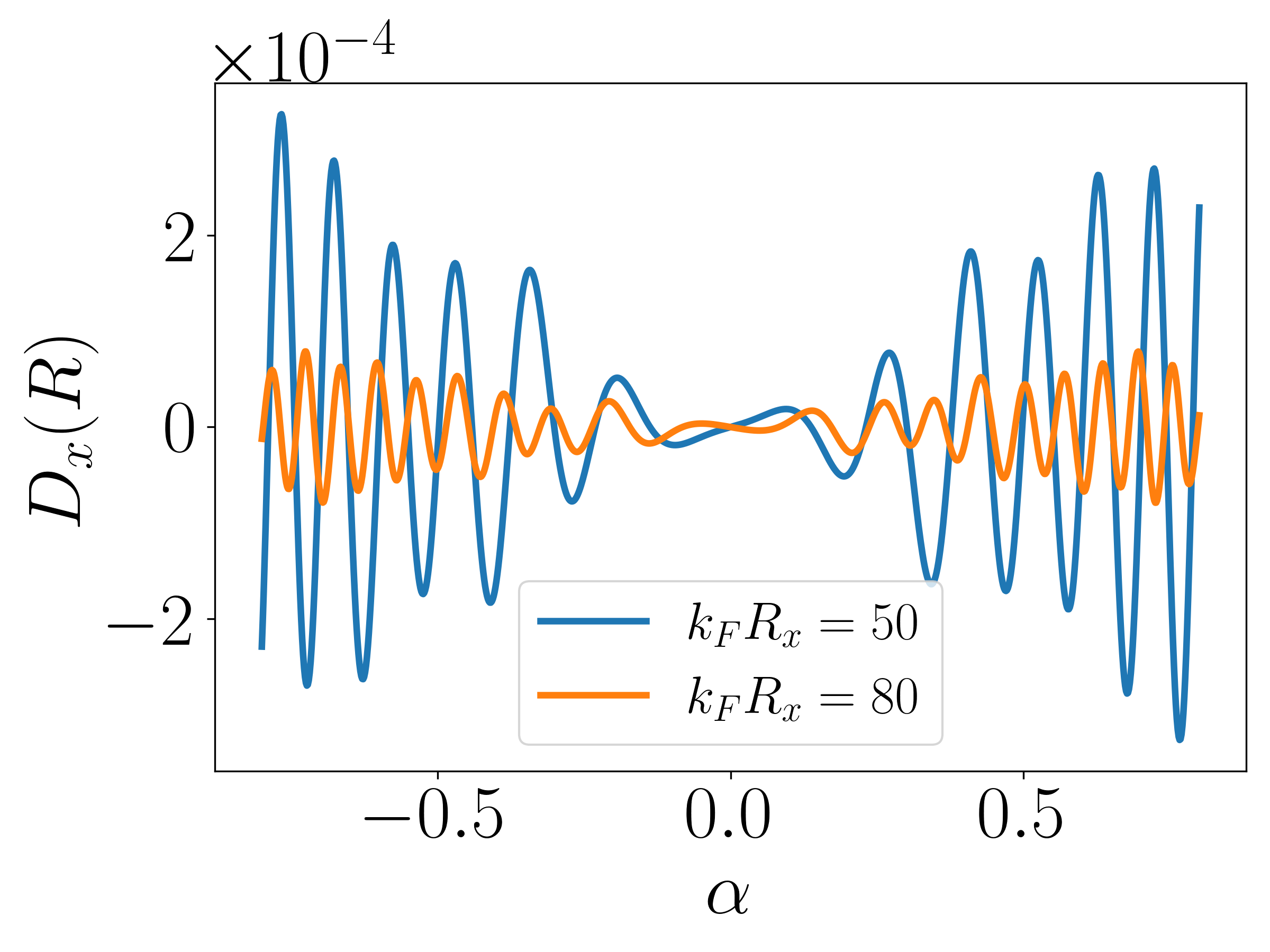}
        \subcaption{}
    \end{minipage}%
    \hfill
    \begin{minipage}[t]{0.45\hsize}
        \centering
        \includegraphics[width=\hsize]{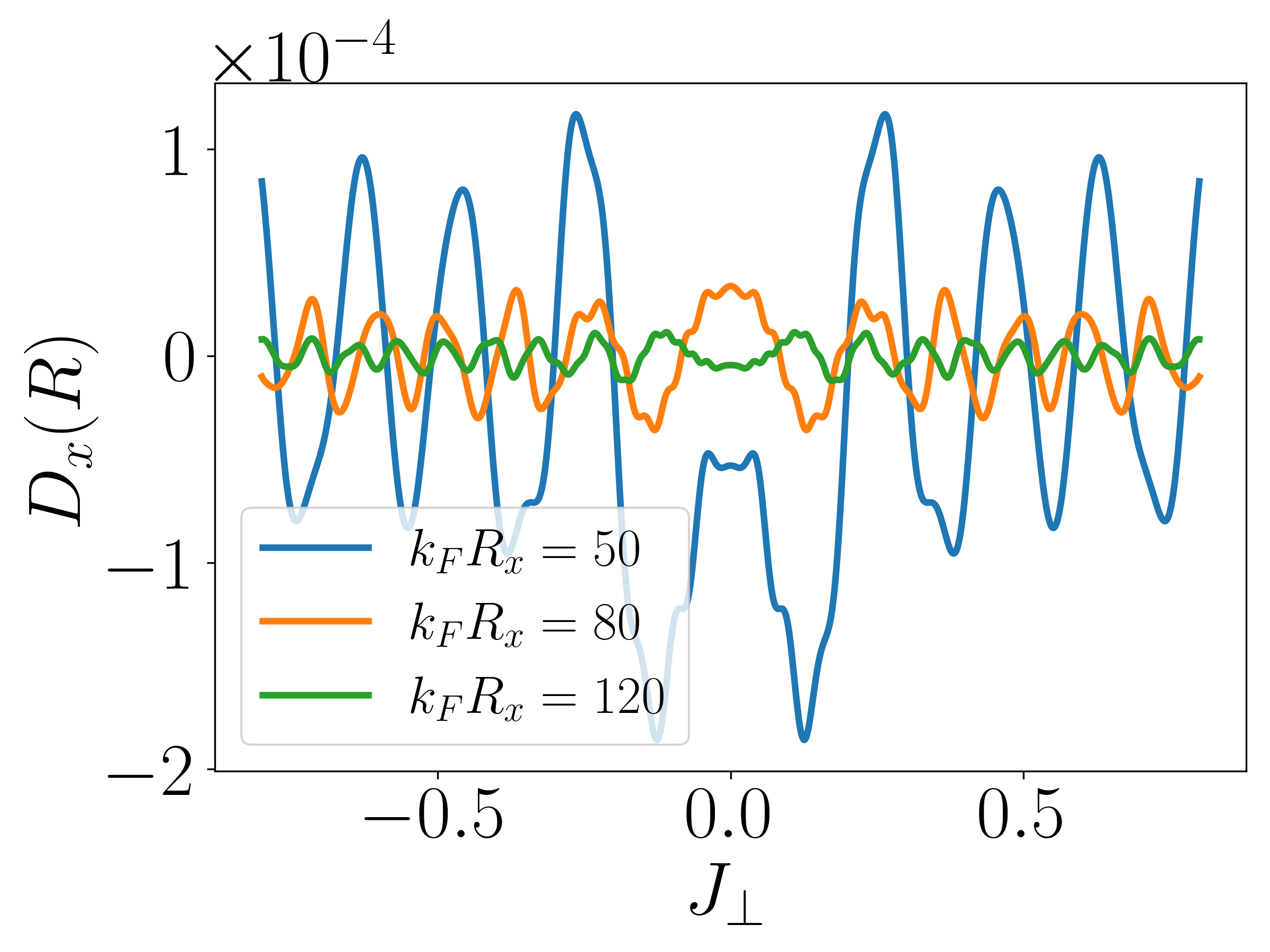}
        \subcaption{}
    \end{minipage}

 
    \begin{minipage}[t]{0.45\hsize}
        \centering
        \includegraphics[width=\hsize, keepaspectratio]{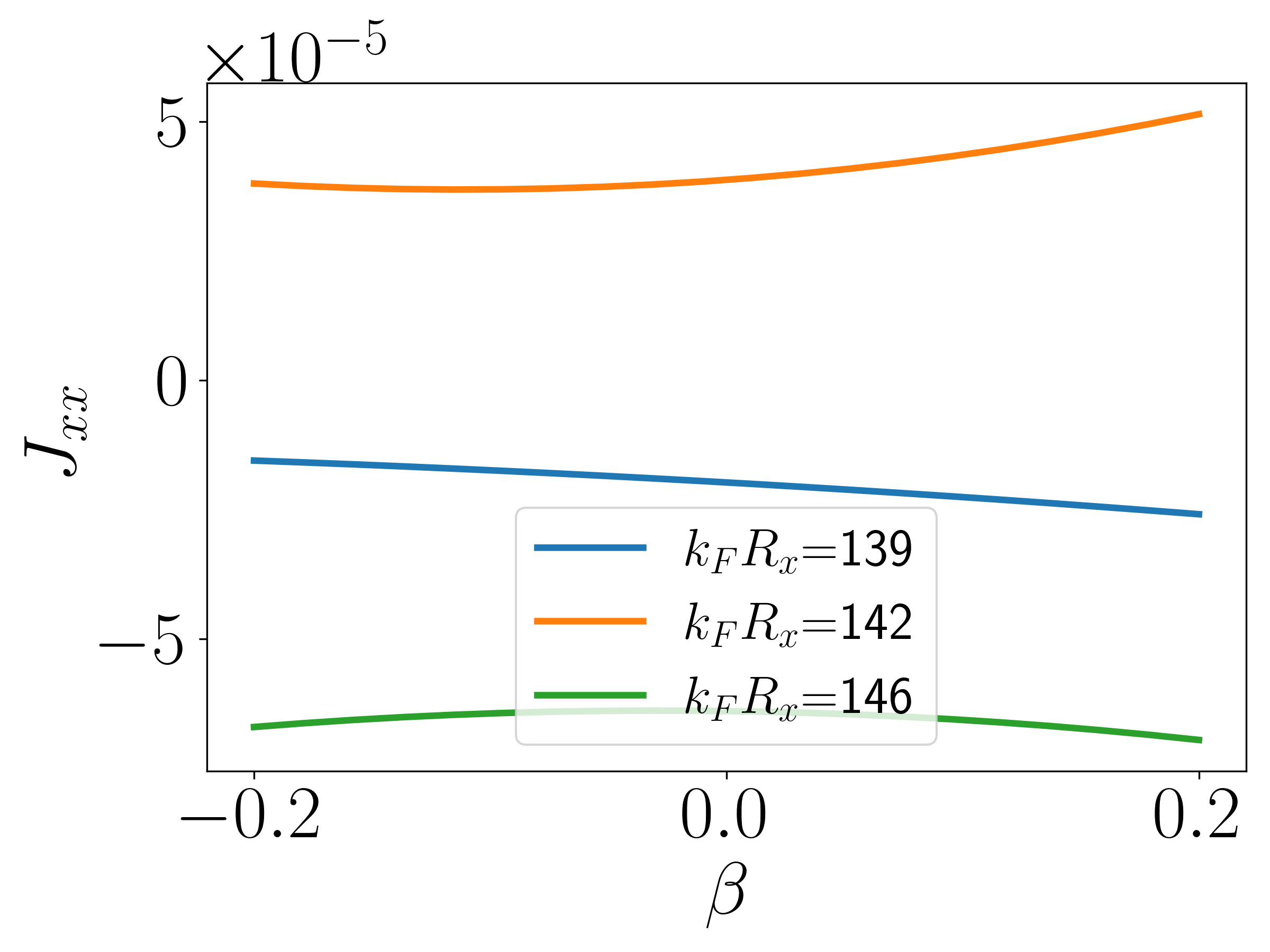}
        \subcaption{}
    \end{minipage}%
    \hfill
    \begin{minipage}[t]{0.45\hsize}
        \centering
        \includegraphics[width=\hsize, keepaspectratio]{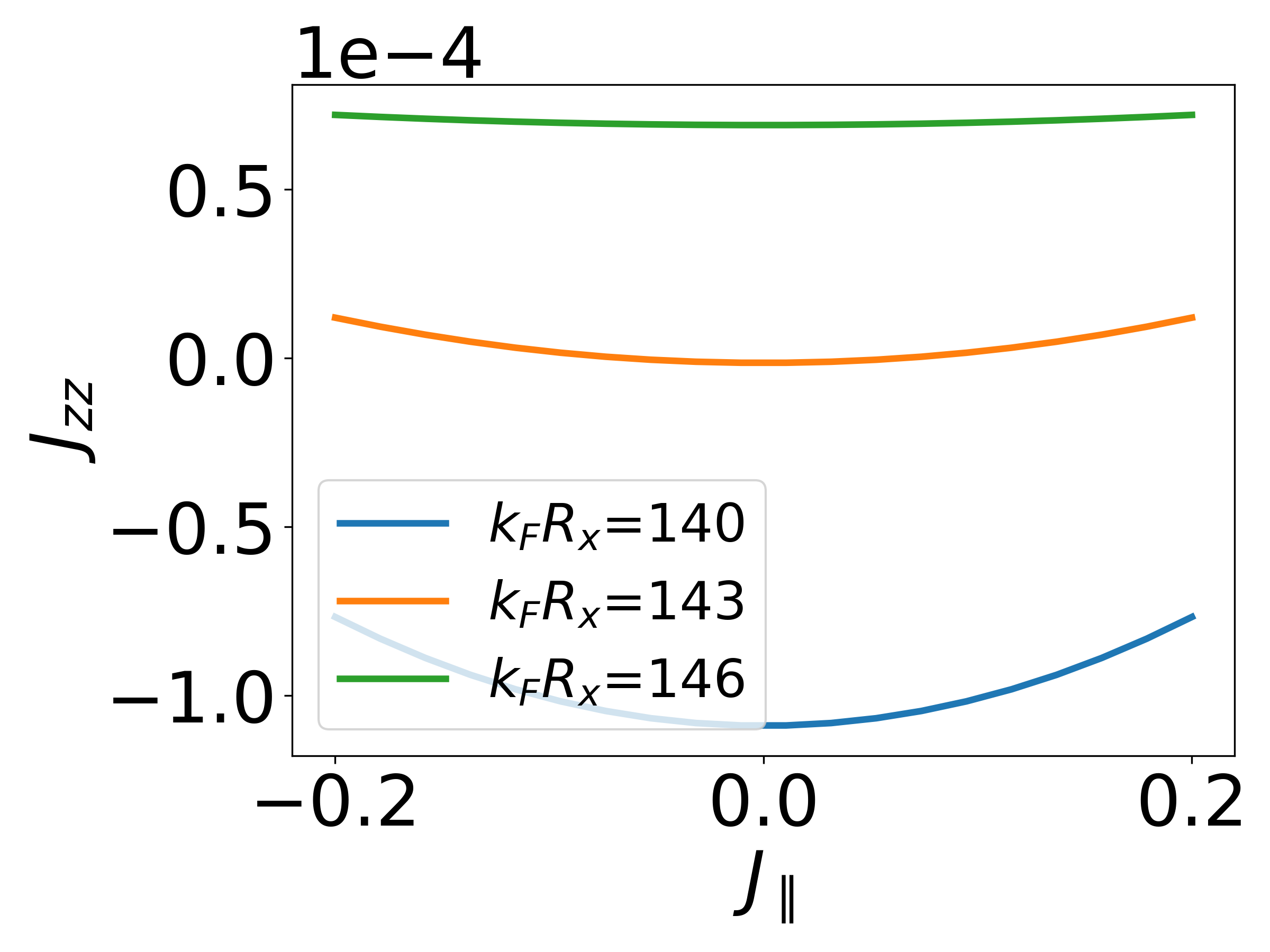}
        \subcaption{}
    \end{minipage}
    
    \caption{(a), (c), Variation of $J_{ii}$ and $D_x$ with  $\alpha$ with specific parameters $\beta /\epsilon_F= 0.1$ and $J_\perp/\epsilon_F = 0.3$. 
(b), (d), variation with $J_\perp$  with specific parameters $\beta/\epsilon_F = 0.1$ and $\alpha /\epsilon_F= 0.3$.
(e),  variation of $J_{xx}$  with $\beta$ with $\alpha /\epsilon_F= 0.3$ and $J_\perp/\epsilon_F= 0.3$.
(f),  $J_{zz}$ vs $J_\parallel$ when $\beta/\epsilon_F = 0.1$ and $\alpha /\epsilon_F= 0.3$. 
For all plots $y$-axis is scaled in arbitrary units (a.u.).
}
\label{fig-alpha-hz-variation}
\end{figure}

Next, we explore the role of the RSOC and the ferromagnetic exchange coupling
strength to the exchange interaction terms.
 Putting $\boldsymbol{m}_{\parallel}=0$,
we obtain analytically, from, Eq. (\ref{eq: gr}) $G_{1(2)}\left(\boldsymbol{R},\epsilon,-\alpha\right)=-G_{1(2)}\left(\boldsymbol{R},\epsilon,\alpha\right)$,
and as a consequence from Eq. (\ref{eq: jii}),  the exchange
terms follow, 
$J_{ii}\left(\boldsymbol{R},\epsilon,-\alpha\right)=J_{ii}\left(\boldsymbol{R},\epsilon,\alpha\right)$
and 
$D_{i}\left(\boldsymbol{R},\epsilon,-\alpha\right)=-D_{i}\left(\boldsymbol{R},\epsilon,\alpha\right)$.
 Similarly, using, $G_{1(2)}\left(\boldsymbol{R},\epsilon,J_{\perp}\right)=G_{1(2)}\left(\boldsymbol{R},\epsilon,J_{\perp}\right)$,
we obtain, $D_{i}\left(\boldsymbol{R},\epsilon,-J_{\perp}\right)=D_{i}\left(\boldsymbol{R},\epsilon,J_{\perp}\right)$.
In short, $J_{ii}$ is even in both $\alpha$ and $J_{\perp}$, however,
$D_{i}$ is odd $\alpha$ and but even in $J_{\perp}$. 
To explore further, in the Fig.~\ref{fig-alpha-hz-variation}, we plot $J_{xx}$ and $D_{x}$ at fixed separation distance
while changing both $\alpha$ and $J_{\perp}$.
The Fig.~\ref{fig-alpha-hz-variation} clearly confirms the properties discussed above. Besides, all the exchange terms show oscillatory behavior with the $\alpha$ and $J_{\perp}$.  

In particular, in Fig.~\ref{fig-alpha-hz-variation}(a), we plot $J_{xx}$ vs $\alpha$ and in Fig.~\ref{fig-alpha-hz-variation}(b), we plot $J_{xx}$ vs $J_\perp$ for two separation distance fixed at $k_FR_x=50$ and $k_FR_x=120$ (for fixed $k_FR_y=15$ ). In Fig.~\ref{fig-alpha-hz-variation}(c), we plot $D_{x}$ vs $\alpha$ and in Fig.~\ref{fig-alpha-hz-variation}(d), we plot $D_{x}$ vs $J_\perp$ for same set of parameters.



The Fig.~\ref{fig-alpha-hz-variation}(e), shows the variation of $J_{xx}$ with
the altermagnetic band parameter $\beta$. The dependence on $\beta$
is strongly nonlinear and asymmetric between positive and negative
values. These nonlinearity and asymmetry are consistently observed
for both $J_{yy}$ and $J_{zz}$ as well. 
In the weak coupling regime, the $D_{x}(\boldsymbol{R})$ and $D_{y}(\boldsymbol{R})$ components of the DM vector are found to be independent of the altermagnetic band parameter $\beta$. This independence arises because the functions $G_0({\bf R},\epsilon)$, $G_1({\bf R},\epsilon)$, $G_2({\bf R},\epsilon)$, and $I(\boldsymbol{R},\epsilon)$ are themselves independent of $\beta$, as evident from Eq. (\ref{eq: jii}).

\subsection{Altermagnetic effects: in presence of a weak in-plane component }
Next, we study the effect of a weak in-plane component of the magnetization vector.
Our model relies on the fact that its validity depends on the 1st order expansion of the in-plane component of magnetization in the expression of Green functions in Eq.~({\ref{eq: gr_k}}). 
We obtain similar spatial variations of the exchange terms in this configuration, which we do not repeat here for brevity.
In Fig. \ref{fig-alpha-hz-variation}(f), we show the variation of the exchange interaction $J_{zz}$ with the $J_\parallel$.
The DM vector we obtain in this limit does not depend on the in-plane magnetization component ${\bf m}_\parallel$ as can be seen from  Eq. (\ref{eq: gr}).

To calculate the exchange interaction for an arbitrary strength of the in-plane ferromagnetic order, an angular term appears in the expression of $D_{\boldsymbol{k}}$ (see SI Eq. (S12)). 
As a consequence,  the analytical calculation of the real-space Green's function becomes very complex.
 To retain analytical clarity and physical insight, which are lost in a fully numerical approach, we restrict our consideration to the weak coupling limit.

\clearpage 
\section{Discussions}
%

We derive the RKKY exchange interaction terms of a 2D ferromagnet with RSOC and altermagnetic bands.
Considering RSOC stronger than altermagnetic band structure, we obtain two key-types of exchange interaction terms, Heisenberg and DM interaction with additional cross terms.
In particular, our model is effective to capture the effect of any strength of RSOC on the exchange interaction terms.
We analyze two ferromagnetic pattern, when moments align along out-of-plane direction and when moments are in-plane. 
For out-of-plane case, the exchange terms decays as $1/R^2$. The Heisenberg and DM-interaction terms both shows beating-like pattern for the spatial variation of the RKKY terms.
We obtain two important symmetries: the Heisenberg terms $J_{ii}$ are always even in RSOC strength $\alpha$ and exchange coupling strength $J_\perp$. 
However, although the DM-terms are even in $J_\perp$ they are odd with the variation of $\alpha$.
Besides, both the Heisenberg and DM terms increse in magnitude with oscillation with the variation of $\alpha$. 
Finally, we show that in the weak limit of altermagnetic band parameter $\beta$, the DM-terms become unaltered with the variation of $\beta$ however, the Heisenberg term shows strong non-linear and asymmetric dependence on the $\beta$.

When the ferromagnetic alignment is along the in-plane direction,  we calculate the RKKY interaction when the exchange coupling strength ($J_\parallel$) between electrons and magnetic moment is small. 
In this limit, the DM interaction is found to be independent of the in-plane ferromagnetic order. In contrast, the Heisenberg exchange exhibits a strong, non-linear dependence on $J_\parallel$, varying evenly with it.

\clearpage   
%
%
%
\clearpage
\onecolumngrid
\appendix
\bibliographystyle{ieeetr}
\bibliography{references_rkky_2025}

\end{document}


\title{
Supplemental Information for 
"RKKY interaction mediated by a spin-polarized 2D electron gas with Rashba and altermagnetic coupling"
}

\author{Anirban Kundu }
\affiliation{Asia Pacific Center for Theoretical Physics, POSTECH, Pohang 37673, Korea.}



\maketitle




%
%
%
\onecolumngrid
\setcounter{section}{0}
\setcounter{figure}{0}
\setcounter{table}{0}
\setcounter{equation}{0}

\renewcommand{\thefigure}{S\arabic{figure}}
\renewcommand{\thetable}{S\arabic{table}}
\renewcommand{\theequation}{S\arabic{equation}}

\section{Angular integration}

The key integral is, 
\begin{equation}
I\left(R,\epsilon\right)=\int d\boldsymbol{k}e^{i\boldsymbol{k}\cdot\boldsymbol{R}}D_{\boldsymbol{k}}^{-1}=\int_{0}^{\infty}dk\frac{k}{\left(\epsilon-\epsilon_{\boldsymbol{k}}+i\delta\right)^{2}-\left(\alpha^{2}k^{2}+J_{\perp}^{2}m_{z}^{2}\right)}\int_{0}^{2\pi}d\phi\text{ }e^{ikR\cos\phi},
\end{equation}
where $|\boldsymbol{R}|=R$. Using the Jacobi--Anger expansion \cite{abramowitz}, 
\begin{equation}
e^{ikR\cos\theta}=\sum_{n=-\infty}^{\infty}i^{n}J_{n}(kR)e^{in\theta}\implies\int_{0}^{2\pi}d\phi\text{ }e^{ikR\cos\phi}=2\pi J_{0}(kR).
\end{equation}
The remaining $k$- integration, 
\begin{equation}
I\left(R,\epsilon\right)=2\pi\int_{0}^{\infty}dk\frac{kJ_{0}(kR)}{\left(\epsilon-\epsilon_{\boldsymbol{k}}+i\delta\right)^{2}-\left(\alpha^{2}k^{2}+J_{\perp}^{2}m_{z}^{2}\right)}=2\pi\int_{-\infty}^{\infty}dk\frac{kH_{0}^{(1)}(kR)}{\left(\epsilon-ak^{2}+i\delta\right)^{2}-\left(\alpha^{2}k^{2}+J_{\perp}^{2}m_{z}^{2}\right)}~,
\end{equation}
where, $a=\hbar^2/2m$. The relevant poles are at, 
\begin{equation}
k_{\pm}=\frac{1}{\sqrt{2}a}\sqrt{\alpha^{2}+2a\epsilon\pm\sqrt{\alpha^{4}+4a\alpha^{2}\epsilon+4a^{2}J_{\perp}^{2}m_{z}^{2}}}+i\delta ~.
\end{equation}
Using, $\left(\epsilon-ak^{2}+i\delta\right)^{2}-\left(\alpha^{2}k^{2}+J_{\perp}^{2}m_{z}^{2}\right)=a^{2}\left(k-k_{+}\right)\left(k-k_{-}\right)\left(k+k_{+}\right)\left(k+k_{-}\right)$
\begin{equation}
I\left(\boldsymbol{R},\epsilon\right)=I\left(R,\epsilon\right)=\frac{\pi}{2a^{2}\left(k_{+}^{2}-k_{-}^{2}\right)}\left(k_{+}H_{0}^{(1)}(k_{+}R)-k_{-}H_{0}^{(1)}(k_{-}R)\right).
\end{equation}
\textit{Asymptotic form}: Since the RKKY exchange is intrinsically
long-ranged, we examine its behaviour in the asymptotic regime $kR\gg1$.
In this limit, the cylindrical Hankel function admits the standard
asymptotic expansion  \cite{arfken}, 
\begin{equation}
H_{0}^{(1)}(kR)\simeq\sqrt{\frac{2}{\pi kR}}\,e^{i(kR-\frac{\pi}{4})},
\end{equation}
which we shall employ to simplify the analytic form of the interaction.
This expression describes an outgoing cylindrical wave in the far-field
region, whose amplitude decays as $1/\sqrt{R}$. Accordingly, the
asymptotic form of the radial integral $I(R,\epsilon)$ for $kR\gg1$
becomes,
\begin{equation}
I\left(R,\epsilon\right)=\frac{1}{a^{2}\left(k_{+}^{2}-k_{-}^{2}\right)}e^{-i\frac{\pi}{4}}\sqrt{\frac{\pi}{2R}}\left(\sqrt{k_{+}}\,e^{ik_{+}R}-\sqrt{k_{-}}\,e^{ik_{-}R}\right), 
\end{equation}
which will be used to evaluate the long-range behaviour of the RKKY
interaction. The above is superposition of two waves with different
frequencies and hence generating beating effects. We will show how
this effect is modified by the RSOC and altermagnetic band parameter.
Small Argument Asymptotic $(\ensuremath{kR\ll1})$: 
\begin{equation}
H_{0}^{(1)}(kR)\sim\frac{2i}{\pi}\left[\ln\left(\frac{kR}{2}\right)+\gamma\right]+\mathcal{O}((kR)^{2}),
\end{equation}
where $\gamma\approx0.5772$ is the \textit{Euler-Mascheroni} constant.
This form shows the logarithmic divergence as $kR\to0$. As a consequence,
\begin{equation}
I\left(R,\epsilon\right)=\frac{i}{a^{2}\left(k_{+}^{2}-k_{-}^{2}\right)}\left(k_{+}\ln\left(\frac{k_{+}R}{2}\right)-k_{-}\ln\left(\frac{k_{-}R}{2}\right)+\gamma\left(k_{+}-k_{-}\right)\right).
\end{equation}

\section{Effect of In-plane component of magnetization}
In the presence of small in-plane magentization, we obtain
\begin{align}
D_{x}\left(\boldsymbol{R}\right) & =-\frac{\mathcal{J}^{2}}{\pi}\text{Im}\int_{-\infty}^{\epsilon_{F}}d\epsilon\text{ }\left(-i\right)\text{ }\left(\epsilon+\frac{\hbar^{2}}{2m}\frac{\partial^{2}}{\partial R^{2}}\right)I\left(R,\epsilon\right)\left(-i2\alpha\frac{\partial}{\partial R_{y}}\right)I\left(R,\epsilon\right),\\
D_{y}\left(\boldsymbol{R}\right) & =-\frac{\mathcal{J}^{2}}{\pi}\text{Im}\int_{-\infty}^{\epsilon_{F}}d\epsilon\text{ }\left(-i\right)\text{ }\left(\epsilon+\frac{\hbar^{2}}{2m}\frac{\partial^{2}}{\partial R^{2}}\right)I\left(R,\epsilon\right)\left(i2\alpha\frac{\partial}{\partial R_{x}}\right)I\left(R,\epsilon\right).
\end{align}
As evident from Eq. (8) the $k_{\pm}\left(\alpha,J_{\perp},\epsilon_{F}\right)$
does not depend on the parameters $m_{x}$, $m_{y}$ and $\beta$.
And as a consequence $I\left(R,\epsilon\right)=I\left(R,\epsilon,\alpha,J_{\perp},\epsilon_{F}\right)$
does not depend on the same set of parameters. And same is true for $D_{x}$ and $D_{y}$.

\section{Green's function  for in-plane ferromagnetic order}

Full Hamiltonian,
\begin{align}
H_{\boldsymbol{k}} & =\epsilon_{\boldsymbol{k}}\sigma_{0}+\left(\alpha\hat{z}\times\boldsymbol{k}+J_{\parallel}\boldsymbol{m}_{\parallel}\right)\cdot\boldsymbol{\sigma}+J_{\perp}\boldsymbol{m}_{\perp}\sigma_{3}+\beta\left(k_{x}^{2}-k_{y}^{2}\right)\sigma_{3}.\label{eq:=000020main_hamiltonian}
\end{align}
where, $\epsilon_{\bm{k}}=\hbar^{2}k^{2}/2m$, $\alpha$ is the Rashba
spin-orbit coupling strength, $\beta$ is the altermagnetic coupling
parameter, $\sigma_{i}$ ($\forall i\in[1,3]$) is the Pauli matrices
and $\sigma_{0}$ is the identity matrix, $\boldsymbol{m}_{\parallel}=m_{x}\hat{x}+m_{y}\hat{y}$
is the in-plane and $\boldsymbol{m}_{\perp}=m_{z}\hat{z}$ is the
out-of-plane component of magnetic moment vector, $J_{\perp}$, $J_{\parallel}$
are the perpendicular and in-plane exchange coupling strength. With
$\beta_{\boldsymbol{k}}=\beta\left(k_{x}^{2}-k_{y}^{2}\right)$, GF
for above Hamiltonain for $J_{\perp}=0$ is, 
\begin{align}
G\left(\boldsymbol{k},\epsilon\right) & =\left(\epsilon-H_{\boldsymbol{k}}+i\eta\right)^{-1}\nonumber\\
 & =D_{\boldsymbol{k}}^{-1}\left[\left(\epsilon-\epsilon_{\boldsymbol{k}}\right)\sigma_{0}+\left(\alpha k_{y}+h_{x}\right)\sigma_{1}+\left(-\alpha k_{x}+h_{y}\right)\sigma_{2}+\beta_{\boldsymbol{k}}\sigma_{3}\right],
\end{align}
where, 
\begin{align}
D_{\boldsymbol{k}}=\left(\epsilon-\epsilon_{\boldsymbol{k}}+i\eta\right)^{2}-\left(\alpha k_{y}+h_{x}\right)^{2}-\left(-\alpha k_{x}+h_{y}\right)^{2}-\beta_{\boldsymbol{k}}^{2}.
\end{align}
Clearly, angle appears in the expression of $D_{\boldsymbol{k}}$.
\bibliographystyle{ieeetr}
\bibliography{supplemental.bib}